\documentclass[preprint, superscriptaddress, preprintnumbers,amsmath,amssymb]{revtex4}

\usepackage{booktabs}
\usepackage{color,graphicx}
\usepackage{amsmath}
\usepackage{dcolumn}
\usepackage{CJK}                 % Align table columns on decimal point
\usepackage{bm}                  % bold math
\usepackage{soul}
\usepackage{enumerate}
\usepackage{multirow}
\usepackage{mathrsfs}
\usepackage{braket}
\usepackage{longtable}
\usepackage{pdflscape}

\usepackage[dvipdfm,
            pdfstartview=FitH,
            CJKbookmarks=true,
            bookmarksnumbered=true,
            bookmarksopen=true,
            colorlinks,
            pdfborder=001,
            linkcolor=blue,
            anchorcolor=blue,
            citecolor=blue
            ]{hyperref}

\usepackage{float}

\begin{document}
\begin{CJK*}{GBK}{song}

\title{Optimized Dirac Woods-Saxon basis for covariant density functional theory}

\author{K. Y. Zhang}
\affiliation{State Key Laboratory of Nuclear Physics and Technology, School of Physics, Peking University, Beijing 100871, China}
\affiliation{Key Laboratory of Neutron Physics, Institute of Nuclear Physics and Chemistry, CAEP, Mianyang, Sichuan, 621900, China}

\author{C. Pan}
\affiliation{State Key Laboratory of Nuclear Physics and Technology, School of Physics, Peking University, Beijing 100871, China}

\author{S. Q. Zhang} \email{sqzhang@pku.edu.cn}
\affiliation{State Key Laboratory of Nuclear Physics and Technology, School of Physics, Peking University, Beijing 100871, China}

\begin{abstract}
  The Woods-Saxon basis has achieved great success in both nonrelativistic and covariant density functional theories in recent years.
  Due to its nonanalytical nature, however, applications of the Woods-Saxon basis are numerically complicated and computationally time consuming.
  In this paper, based on the deformed relativistic Hartree-Bogoliubov theory in continuum (DRHBc), we check in detail the convergence with respect to the basis space in the Dirac sea.
  An optimized Dirac Woods-Saxon basis is proposed, whose corresponding potential is close to the nuclear mean field.
  It is shown that the basis space of the optimized Dirac Woods-Saxon basis required for convergence is substantially reduced compared with the original one.
  In particular, it does not need to contain the bases from continuum in the Dirac sea.
  The application of the optimized Woods-Saxon basis would greatly reduce computing resource for large-scale density functional calculations.
\end{abstract}

\date{\today}

\maketitle

%%%%%%%%%%%%%%%%%%%%%%%%%%%%%%%%%%%%%%%%%%%%%%%%%%%%%%%%%%
%                    begin  introduction
%%%%%%%%%%%%%%%%%%%%%%%%%%%%%%%%%%%%%%%%%%%%%%%%%%%%%%%%%%

%----------------------------------------------------------------------------------------
\section{Introduction}
%----------------------------------------------------------------------------------------

The study of exotic nuclei far from the $\beta$ stability line has been at the forefront of nuclear physics research since the discovery of the first halo nucleus $^{11}$Li in the 1980s~\cite{Tanihata1985PRL}.
In addition to halos, novel phenomena found in exotic nuclei include the pygmy resonance~\cite{Adrich2005PRL}, the disappearance of traditional magic numbers and the emergence of new ones~\cite{Ozawa2000PRL}.
These phenomena not only promote the worldwide development of radioactive ion beam facilities but also challenge conventional nuclear models.

The nuclear density functional theory (DFT) is able to provide a unified description for almost all nuclei in the nuclear chart and has become one of the most important microscopic methods for the study of nuclear structure~\cite{Bender2003RMP}.
Its relativistic version, i.e., the covariant density functional theory (CDFT), has attracted wide attention in recent years for many advantages, such as the automatic inclusion of the nucleonic spin degree of freedom and the spin-orbital interaction~\cite{Ren2020PRC(R)}, the explanation of the pseudospin symmetry in the nucleon spectrum~\cite{Ginocchio1997PRL,Meng1998Phys.Rev.C628,Meng1999PRC,Chen2003CPL,Ginocchio2005PhysRep,Liang2015PhysRep} and the spin symmetry in the antinucleon spectrum~\cite{Zhou2003PRL,He2006EPJA,Liang2015PhysRep}, and the natural inclusion of the nuclear magnetism~\cite{Koepf1989NPA}, which plays an important role in nuclear magnetic moments~\cite{Yao2006Phys.Rev.C24307,Arima2011,Li2011Sci.ChinaPhys.Mech.Astron.204,Li2011Prog.Theor.Phys.1185,Li2018Front.Phys.Beijing132109} and nuclear rotations~\cite{Meng2013FOP,Konig1993PRL,Afanasjev2000NPA,PhysRevC.62.031302,PhysRevC.82.034329,Zhao2011PRL,Zhao2011PLB,Zhao2012Phys.Rev.C54310,Zhao2015PRL,Wang2017Phys.Rev.C54324,Wang2018Phys.Rev.64321,Ren2019SciChina,Ren2020NPA}.

In the nuclear DFT, the equations of motion for (quasi)nucleons are either solved in coordinate space or transformed into a matrix diagonalization problem in a complete basis, e.g., the harmonic oscillator (HO) basis.
As a good approximation for the mean field of stable nuclei, the HO potential can be solved analytically, and the HO wave function in analytical form may bring convenience in the calculation of some matrix elements, e.g., the separable pairing force~\cite{Tian2009PLB}.
Due to the incorrect asymptotic behavior of HO wave functions, however, the expansion in a localized HO basis is incapable of describing nuclei with very diffuse spatial density distributions.
To improve the asymptotic behavior of HO wave functions, a transformed HO basis has been proposed in Refs.~\cite{Stoitsov1998PRC(1),Stoitsov1998PRC(2)} via a local scaling transformation.

Solutions in coordinate space can describe properly the asymptotic behavior of wave functions.
When dealing with systems having small separation energy, the coordinate-space calculations in large boxes were found to be more effective than the transformed HO basis~\cite{Zhang2013PRC}.
In the CDFT, the relativistic Hartree equation that neglects pairing correlations has been successfully solved on a three-dimensional lattice~\cite{Ren2017PRC}.
For the treatment of pairing correlations in weakly bound nuclei, it was shown that the Bogoliubov transformation is more suitable than the widely used Bardeen-Cooper-Schrieffer (BCS) method~\cite{Dobaczewski1984NPA,Meng1998NPA}.
However, the relativistic Hartree-Bogoliubov equation in coordinate space has only been solved by assuming spherical symmetry~\cite{Meng1996PRL,Poschl1997PRL,Meng1998PRL}.
Solving the deformed relativistic Hartree-Bogoliubov equation in coordinate space is extremely difficult if not impossible~\cite{Zhou2000CPL}.

A Woods-Saxon basis was proposed in Ref.~\cite{Zhou2003PRC} as a reconciler between the HO basis and coordinate space.
It is obtained by solving the Schr\"{o}dinger equation or the Dirac equation containing spherical Woods-Saxon potentials with box boundary conditions, and is correspondingly referred to as the Schr\"{o}dinger Woods-Saxon (SWS) or Dirac Woods-Saxon (DWS) basis.
The Woods-Saxon basis has the advantage in providing appropriate asymptotic behaviors of wave functions because of the nature of the Woods-Saxon type potential~\cite{WS1954PR}.
It was shown in Ref.~\cite{Zhou2003PRC} that for spherical systems the solution of the relativistic Hartree equations in the Woods-Saxon basis is almost equivalent to the solution in coordinate space.
Up to the present, the SWS basis has been applied to the spherical Hartree-Fock-Bogoliubov theory~\cite{Schunck2008PRC(R),Schunck2008PRC}, and the DWS basis to the spherical relativistic Hartree theory~\cite{Zhou2003PRC}, spherical relativistic Hartree-Fock-Bogoliubov theory~\cite{Long2010PRC,Long2010PRC(R)}, deformed relativistic Hartree-Bogoliubov theory in continuum (DRHBc)~\cite{Zhou2010PRC(R),Li2012PRC}, deformed relativistic Hartree-Fock theory~\cite{Geng2020PRC}, and deformed relativistic Hartree-Fock-Bogoliubov theory~\cite{Geng2022PRC}.

Recently, the DRHBc theory has been successful in
studying deformed halo nuclei~\cite{Zhou2010PRC(R),Li2012PRC,Sun2018PLB,Zhang2019PRC,Sun2020NPA,Yang2021PRL,Sun2021PRC(1),Zhong2022SciChina},
investigating the deformation effects on the location of the neutron drip line~\cite{In2021IJMPE},
predicting stability peninsulas beyond the primary neutron drip line~\cite{Zhang2021PRC(L),Pan2021PRC,He2021CPC},
exploring rotational excitations of exotic nuclei with the angular momentum projection~\cite{Sun2021SciBull,Sun2021PRC(2)},
and revealing the shape coexistence from light to heavy nuclei~\cite{In2020JKPS,Choi2022PRC,Kim2022PRC}.
In particular, stimulated by the success of the fist nuclear mass table with continuum effects~\cite{Xia2018ADNDT},
many efforts have been made to construct an upgraded mass table including simultaneously deformation and continuum effects based on the DRHBc theory~\cite{Zhang2020PRC},
and the DRHBc mass table for even-even nuclei has just been published~\cite{Zhang2022ADNDT}.
The next step of the DRHBc mass table is the systematical calculation of odd-mass and odd-odd nuclei in the nuclear chart~\cite{Pan2022PRC}.
As mentioned in Ref.~\cite{Zhang2020PRC}, since the DRHBc theory with the DWS basis is numerically very complicated, large-scale DRHBc calculations are extremely time consuming, especially for odd nuclei where the blocking effects should be considered.

This paper is devoted to exploring methods to reduce the computing cost of the DRHBc theory from the perspective of the DWS basis space.
On the one hand, the DWS basis space includes both bases in the Fermi sea and in the Dirac sea for completeness~\cite{Zhou2003PRC}.
In order to guarantee the convergence with respect to the basis space, it was suggested to take the number of DWS bases in the Dirac sea as the same as that in the Fermi sea~\cite{Zhou2010PRC(R),Li2012PRC}, and later all DRHBc studies followed this way.
However, in the spherical relativistic Hartree-Fock-Bogoliubov~\cite{Long2010PRC,Long2010PRC(R)} and deformed relativistic Hartree-Fock~\cite{Geng2020PRC} calculations, the number of DWS bases in the Dirac sea is only about one third of that in the Fermi sea.
On the other hand, usually the Woods-Saxon potential for the basis is parameterized~\cite{Koepf1991ZPA}.
It is natural to consider the possibility of an optimized DWS basis whose corresponding potential is closer to the mean field of the calculated nucleus, and as a result the basis space required for convergence would be reduced.
Following this idea we examine in detail the convergence behavior against the basis truncation in the Dirac sea and propose an optimized Woods-Saxon basis for the large-scale DRHBc calculations as well as other DFT calculations in the present paper.

This paper is organized as follows: In Sec.~\ref{theory}, we introduce the DRHBc theory and the DWS basis.
The numerical details are given in Sec.~\ref{numerical}.
The results and discussion are presented in Sec.~\ref{results}.
Finally, a summary is given in Sec.~\ref{summary}.

%%%%%%%%%%%%%%%%%%%%%%%%%%%%%%%%%%%%%%%%%%%%%%%%%%%%%%%%%%
%                    begin  theoretical framework
%%%%%%%%%%%%%%%%%%%%%%%%%%%%%%%%%%%%%%%%%%%%%%%%%%%%%%%%%%

%----------------------------------------------------------------------------------------
\section{Theoretical framework}\label{theory}
%----------------------------------------------------------------------------------------

The details of the DRHBc theory with meson-exchange and point-coupling density functionals can be found in Refs.~\cite{Li2012PRC} and \cite{Zhang2020PRC} respectively. In the DRHBc theory, the relativistic Hartree-Bogoliubov (RHB) equation reads~\cite{Kucharek1991ZPA}
\begin{equation}\label{RHB}
\left(\begin{matrix}
h_D-\lambda & \Delta \\
-\Delta^* &-h_D^*+\lambda
\end{matrix}\right)\left(\begin{matrix}
U_k\\
V_k
\end{matrix}\right)=E_k\left(\begin{matrix}
U_k\\
V_k
\end{matrix}\right),
\end{equation}
where $\lambda$ is the Fermi energy, and $E_k$ and $(U_k, V_k)^{\rm T}$ are the quasiparticle energy and wave function, respectively.
$h_D$ is the Dirac Hamiltonian,
\begin{equation}\label{diracH}
h_D(\bm{r})=\bm{\alpha}\cdot\bm{p}+V(\bm{r})+\beta[M+S(\bm{r})].
\end{equation}
$\Delta$ is the pairing potential,
\begin{equation}\label{Delta}
\Delta(\bm r_1,\bm r_2) = V^{\mathrm{pp}}(\bm r_1,\bm r_2)\kappa(\bm r_1,\bm r_2),
\end{equation}
with a density-dependent force of zero range,
\begin{equation}\label{pair}
V^{\mathrm{pp}}(\bm r_1,\bm r_2)= V_0 \frac{1}{2}(1-P^\sigma)\delta(\bm r_1-\bm r_2)\left(1-\frac{\rho(\bm r_1)}{\rho_{\mathrm{sat}}}\right),
\end{equation}
and the pairing tensor $\kappa(\bm r_1,\bm r_2)$~\cite{Peter1980Book}.
In principle, the zero-range pairing force is not naturally converged with the pairing window,
and both appropriate pairing strength and pairing window are significant in the present DRHBc calculations~\cite{Zhang2020PRC}.
It is expected to implement the finite-range pairing force, e.g., the Gogny~\cite{Meng1998NPA} or separable pairing force~\cite{Tian2009PLB},
in the DRHBc theory in future work.

In the DRHBc theory, the scalar potential $S(\bm{r})$ and vector potential $V(\bm{r})$ in Eq.~\eqref{diracH} are expanded in terms of the Legendre polynomials,
\begin{equation}\label{legendre}
f(\bm r)=\sum_\lambda f_\lambda(r)P_\lambda(\cos\theta),~~\lambda=0,2,4,\cdots;
\end{equation}
so are the pairing potential and various densities.
In order to properly describe the large spatial extension of weakly bound nuclei, the RHB equation, (\ref{RHB}), is solved in the DWS basis~\cite{Zhou2003PRC}.
After solving the RHB equations self-consistently, the total energy, rms radii, quadrupole deformation, and other physical quantities can be calculated.

The wave function of the DWS basis can be written as
\begin{equation}
\phi_{n\kappa m}(\bm r s p)=i^p \frac{R_{n\kappa}(r,p)}{r}\mathcal{Y}_{\kappa m}^{l(p)}(\Omega,s),
\end{equation}
where $n$, $\kappa$, and $m$ are its quantum numbers, which will be introduced below.
$\bm r$, $s$, and $p$ are spatial coordinate, spin, and index ($p=1$ or $2$) for upper or lower component.
$R_{n\kappa}$ is the radial wave function,
\begin{equation}
R_{n\kappa}(r,1)=G_{n,\kappa}(r),~~R_{n\kappa}(r,2)=F_{n,\kappa}(r),
\end{equation}
satisfying the radial Dirac equations
\begin{equation}
\begin{aligned}
(-\frac{\partial}{\partial r}+\frac{\kappa}{r})F_{n,\kappa} + [V_{\mathrm{WS}}(r) + S_{\mathrm{WS}}(r) + M] G_{n,\kappa} & = \epsilon_n G_{n,\kappa}, \\
(+\frac{\partial}{\partial r}+\frac{\kappa}{r})G_{n,\kappa} + [V_{\mathrm{WS}}(r) - S_{\mathrm{WS}}(r) - M] F_{n,\kappa} & = \epsilon_n F_{n,\kappa},
\end{aligned}
\label{rDirac}
\end{equation}
where $\epsilon_n$ is the eigenvalue (energy), and $n$ is the node number.
$V_{\mathrm{WS}}+S_{\mathrm{WS}}$ and $V_{\mathrm{WS}}-S_{\mathrm{WS}}$ are parameterized Woods-Saxon potentials for the basis,
\begin{equation}
\begin{aligned}
V_{\mathrm{WS}}+S_{\mathrm{WS}} = \frac{V^+}{1+e^{(r-R^+)/a^+}},\\
V_{\mathrm{WS}}-S_{\mathrm{WS}} = \frac{V^-}{1+e^{(r-R^-)/a^-}},
\end{aligned}
\end{equation}
where $V^\pm$, $R^\pm$, and $a^\pm$ depict the depth, width, and diffuseness of the potential, respectively.
A usual choice of the parametrization can be found in Ref.~\cite{Koepf1991ZPA}.
$\mathcal{Y}_{\kappa m}^{l}$ is the spinor spherical harmonic,
\begin{equation}
\mathcal{Y}_{\kappa m}^l(\Omega,s)=\sum_{m_l,m_s} \langle\frac{1}{2}m_s l m_l| j m\rangle Y_{l m_l}(\Omega)\chi_{\frac{1}{2}m_s}(s),
\end{equation}
where $Y_{l m_l}$ is the spherical harmonic function, $\chi_{\frac{1}{2}m_s}$ is the spin wave function, $l$ is the orbital angular momentum, $j$ is the total angular momentum, and $m_l$, $m_s$, and $m$ are the third components of $l$, $s$, and $j$, respectively.
The quantum number $\kappa$ is given by the parity $\pi$ and $j$, $\kappa=\pi(-1)^{j+1/2}(j+1/2)$, running over positive and negative integers $\pm1, \pm2, \cdots$.
The orbital angular momenta for upper and lower components are
\begin{equation}
l(1)=j+\frac{1}{2}\text{sgn}(\kappa),~~l(2)=j-\frac{1}{2}\text{sgn}(\kappa),
\end{equation}
respectively.

%%%%%%%%%%%%%%%%%%%%%%%%%%%%%%%%%%%%%%%%%%%%%%%%%%%%%%%%%%
%                    begin  numerical details
%%%%%%%%%%%%%%%%%%%%%%%%%%%%%%%%%%%%%%%%%%%%%%%%%%%%%%%%%%

%----------------------------------------------------------------------------------------
\section{Numerical details}\label{numerical}
%----------------------------------------------------------------------------------------

Taking light $^{20}$Ne, medium-heavy $^{112}$Mo, and heavy $^{300}$Th nuclei as examples, a detailed convergence check of the total energy against the energy cutoff in the Dirac sea $E_{\mathrm{cut}}^D $ will be performed.
The radial Dirac equations for the DWS basis, \eqref{rDirac}, are solved with a box size $R_{\mathrm{box}}=20$ fm and a mesh size $\Delta r = 0.1$ fm~\cite{Zhou2003PRC,Zhang2020PRC}.
The angular momentum cutoff for the DWS basis is $J_{\mathrm{max}} = \frac{23}{2}~\hbar$~\cite{Zhang2020PRC}.
The Legendre expansion truncations in Eq.~\eqref{legendre} are chosen as $\lambda_{\mathrm{max}} = 6$ and $8$ for nuclei with $8 \le Z \le 70$ and $72 \le Z \le
100$, respectively~\cite{Pan2019IJMPE,Zhang2020PRC}.
The above numerical details are the same as those used in the DRHBc mass table calculations for even-even nuclei~\cite{Zhang2022ADNDT}.
Since a zero-range pairing interaction \eqref{pair} is adopted in the DRHBc theory, in the numerical check of the convergence with respect to the basis space pairing correlations should be neglected.
In the DRHBc mass table calculations~\cite{Zhang2022ADNDT}, the energy cutoff for the DWS basis in the Fermi sea $E_{\mathrm{cut}}^F = 300$ MeV and the number of DWS bases in the Dirac sea is the same as that in the Fermi sea, which are able to give converged results as shown in Ref.~\cite{Zhang2020PRC} and thus are taken as the benchmark in the following.

%%%%%%%%%%%%%%%%%%%%%%%%%%%%%%%%%%%%%%%%%%%%%%%%%%%%%%%%%%
%                    begin  results and discussions
%%%%%%%%%%%%%%%%%%%%%%%%%%%%%%%%%%%%%%%%%%%%%%%%%%%%%%%%%%

%----------------------------------------------------------------------------------------
\section{Results and discussion}\label{results}

%----------------------------------------------------------------------------------------
\begin{figure}[htbp]
  \centering
  \includegraphics[width=0.4\textwidth]{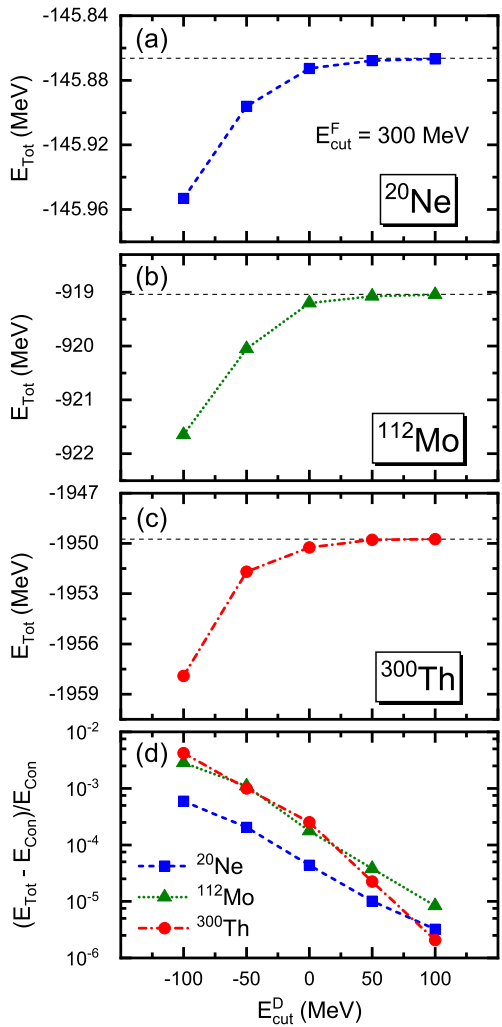}
  \caption{The total energy is shown as a function of the energy cutoff in the Dirac sea $E_{\mathrm{cut}}^D$ for $^{20}$Ne (a), $^{112}$Mo (b), and $^{300}$Th (c). The horizontal dashed lines show their total energies, calculated with the numerical condition that the number of DWS bases in the Dirac sea is the same as that in the Fermi sea, and are regarded as the converged results. In (d), the relative difference of the total energy from the converged one, $(E_{\mathrm{Tot}}-E_{\mathrm{Con}})/E_{\mathrm{Con}}$, is plotted as a function of $E_{\mathrm{cut}}^D$ for $^{20}$Ne, $^{112}$Mo, and $^{300}$Th. The energy cutoff in the Fermi sea $E_{\mathrm{cut}}^F$ is taken as 300 MeV.}
\label{fig1}
\end{figure}
%----------------------------------------------------------------------------------------

Figure \ref{fig1} shows the convergence of the total energy with respect to the energy cutoff in the Dirac sea $E_{\mathrm{cut}}^D$ (whose zero point is set to be the continuum threshold in the Dirac sea) for $^{20}$Ne, $^{112}$Mo, and $^{300}$Th.
From panels (a), (b), and (c) one finds that, although the total energies of the three nuclei are very different, they show a similar convergence trend with the increase of $E_{\mathrm{cut}}^D$.
When $E_{\mathrm{cut}}^D = 50$ MeV, the difference of the calculated total energy from the converged one (horizontal dashed line) becomes marginal in their respective scale.
It is clearly seen in panel (d) that their relative differences from the converged results are on the order of $10^{-5}$.
Therefore, $E_{\mathrm{cut}}^D = 50$ MeV for the DWS basis can lead to a satisfactory accuracy in the DRHBc calculation.
The $E_{\mathrm{cut}}^D$ value greater than zero suggests that the contribution of the bases from continuum in the Dirac sea is non-negligible with the adopted DWS basis.
A similar conclusion was found in a recent study based on the axially deformed relativistic Hartree-Fock-Bogoliubov theory with the density functional PKA1~\cite{Geng2022PRC}.

%----------------------------------------------------------------------------------------
\begin{figure}[htbp]
  \centering
  \includegraphics[width=0.4\textwidth]{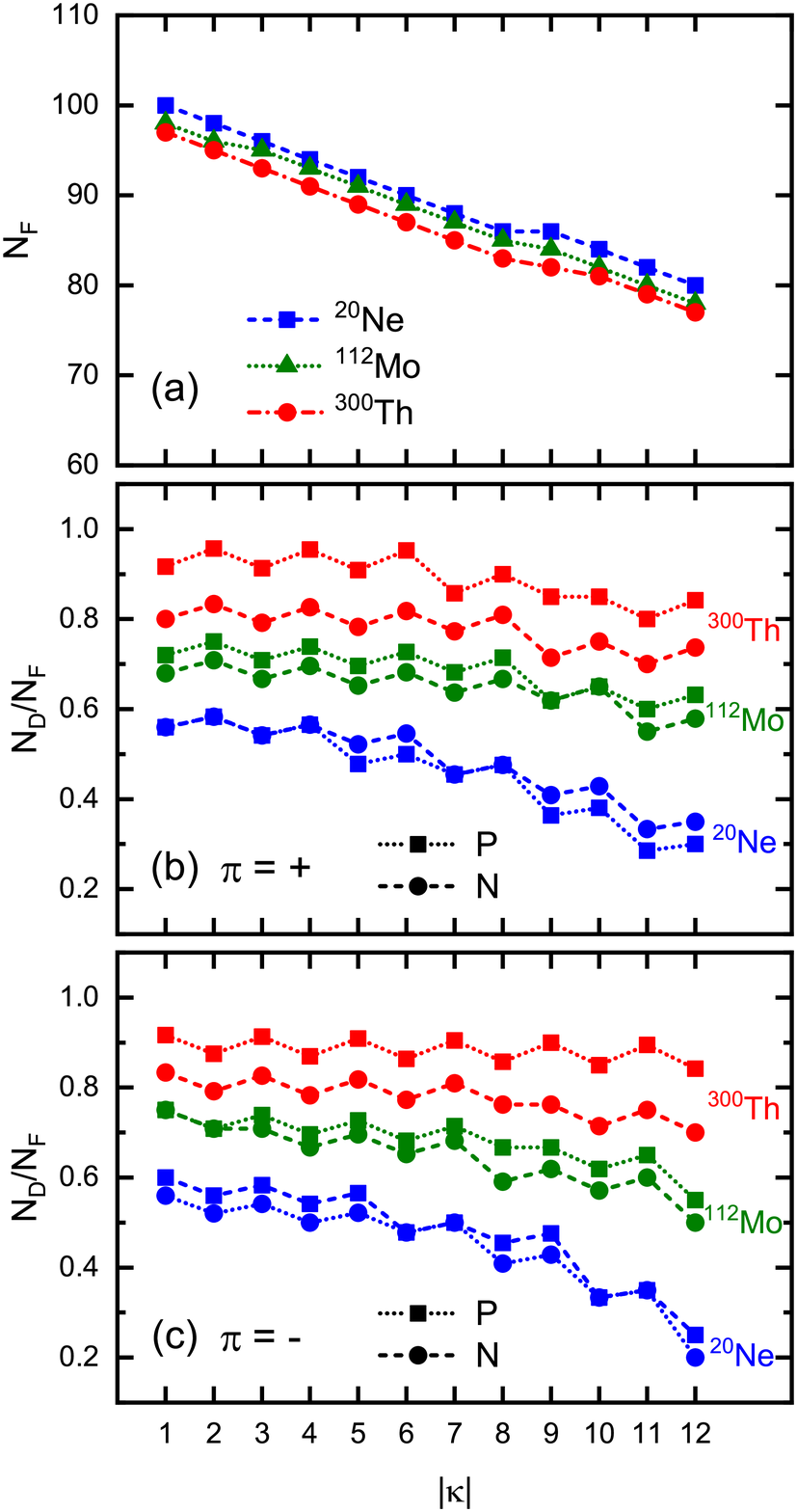}
  \caption{The number of DWS bases in the Fermi sea, $N_F$, and the ratio of the number of DWS bases in the Dirac sea to it, $N_D/N_F$, are shown as functions of the absolute value of the quantum number $\kappa$ in the calculation of $^{20}$Ne, $^{112}$Mo, and $^{300}$Th. For clearness, the ratios for positive parity ($+$) and negative parity ($-$) are shown respectively in (b) and (c), where the values for protons ($P$) and neutrons ($N$) are distinguished by symbols. The energy cutoff in the Fermi sea $E_{\mathrm{cut}}^F$ and that in the Dirac sea $E_{\mathrm{cut}}^D$ are taken as 300 and 50 MeV, respectively.}
\label{fig2}
\end{figure}
%----------------------------------------------------------------------------------------

To compare the number of bases in the Dirac sea $N_D$ with $E_{\mathrm{cut}}^D = 50$ MeV and that in the Fermi sea $N_F$ with $E_{\mathrm{cut}}^F = 300$ MeV, Fig.~\ref{fig2} exhibits $N_F$ and the ratio $N_D/N_F$ as functions of the absolute value of the quantum number $\kappa$ in the calculations of $^{20}$Ne, $^{112}$Mo, and $^{300}$Th.
The $N_F$ as a function of $|\kappa|$ is shown in Fig.~\ref{fig2}(a), where the basis numbers for neutrons/protons and positive/negative parities are close to each other and summed.
First, we can see that the $N_F$s are not very different for $^{20}$Ne, $^{112}$Mo, and $^{300}$Th because of the large energy cutoff $E_{\mathrm{cut}}^F = 300$ MeV, and they decrease slowly and almost linearly with $|\kappa|$.
Second, as shown in Figs.~\ref{fig2}(b) and \ref{fig2}(c), all the ratios of $N_D/N_F$ are smaller than 1, meaning that the basis space in the Dirac sea is not necessarily as large as that in the Fermi sea in order to obtain converged results.
Third, the ratio of $N_D/N_F$ shows an odd-even staggering with the increase of $|\kappa|$.
This is because, for the same $|\kappa|$, the orbital angular momentum $l$ of the basis in the Fermi sea and that in the Dirac sea differs by $1$, e.g., for positive parity the ratio is calculated in the sequence of $s_{1/2}/p_{1/2}$, $d_{3/2}/p_{3/2}$, $d_{5/2}/f_{5/2}$, $\cdots$ for $|\kappa| = 1, 2, 3, \cdots$.
In a basis space truncated by energy, the number of bases corresponding to spin-orbit partners is in general the same, and it decreases with the increasing $l$.
As a result, the numerator $N_D$ and denominator $N_F$ do not decrease synchronously, leading to the odd-even staggering.
Fourth, the ratio decreases with $|\kappa|$ on the whole, which reflects the influence of the different depths of the potentials in the Fermi sea and Dirac sea.
Finally, the ratios of $N_D/N_F$ for the heavy nucleus $^{300}$Th are larger than those for lighter ones $^{20}$Ne and $^{112}$Mo, which shows that the bases in the Dirac sea are more important for heavy nuclei and is consistent with the finding in Ref.~\cite{Geng2022PRC}.
It is also noted that the ratios for protons are obviously larger than those for neutrons in the calculation of $^{300}$Th, which will be further discussed below.

%----------------------------------------------------------------------------------------
\begin{figure}[htbp]
  \centering
  \includegraphics[width=0.4\textwidth]{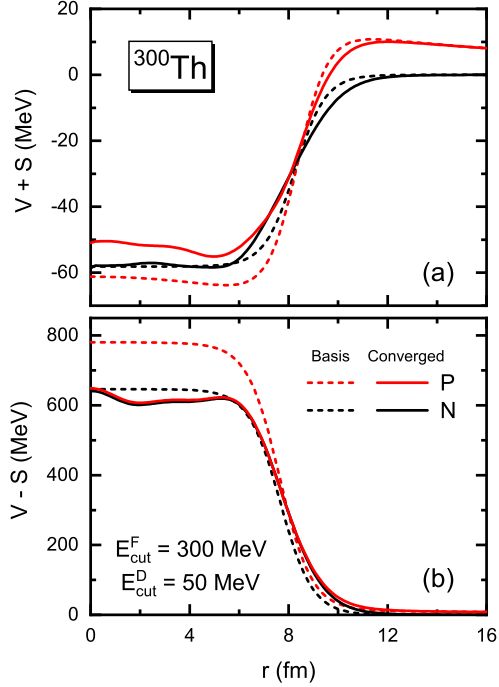}
  \caption{The Woods-Saxon potentials $V+S$ (a) and $V-S$ (b) used to generate the DWS basis and those corresponding to the converged mean field are plotted as functions of the radial coordinate $r$ for $^{300}$Th. Here the angular dependence of the deformed mean-field potentials has been averaged. The energy cutoff in the Fermi sea $E_{\mathrm{cut}}^F$ and that in the Dirac sea $E_{\mathrm{cut}}^D$ are taken as 300 and 50 MeV, respectively.}
\label{fig3}
\end{figure}
%----------------------------------------------------------------------------------------

Is it possible to optimize the usual DWS basis to achieve a better performance?
We note that Ref.~\cite{Zhou2003PRC} has found when the Woods-Saxon potentials for the basis become more different from the converged mean-field potentials of the calculated nucleus, the contribution of the basis in the Dirac sea will be larger.
To illustrate this difference in potentials, Fig.~\ref{fig3} compares for $^{300}$Th the Woods-Saxon potentials for the basis and the converged mean-field ones, and for the latter the angular dependence has been averaged.
For the $V+S$ potential of neutrons, the difference mainly appears in 6 $\le r \le$ 12 fm and is not remarkable.
For the $V-S$ potential of neutrons, distinguishable difference can also be found in 0 $< r \le$ 6 fm.
For protons, however, one finds significant differences in both $V+S$ and $V-S$ potentials.
The depth of the $V+S$ potential differs by about 10 MeV, and the difference in that of the $V-S$ potential is larger than 100 MeV.
This explains why the ratios $N_D/N_F$ for protons are obviously larger than those for neutrons in Fig.~\ref{fig2}(c).

According to the above discussion, a possible way to optimize the DWS basis and reduce the basis space is using the Woods-Saxon potentials close to the converged mean-field ones to generate a new basis replacing the original DWS basis.
As an attempt, for deformed nuclei, we solve a Dirac equation containing the angle averaged converged mean-field potentials to obtain the new basis, referred to as the optimized Dirac Woods-Saxon (ODWS) basis.
This optimized basis is in spirit somewhat similar to the method proposed in Refs.~\cite{Zhang1988PLB,Zhang1991NPA}, where deformed relativistic Hartree equations are solved by expanding the Dirac spinors in terms of basis functions calculated from the self-consistent spherical Hartree potential of the same or a nearby nucleus.
At the end of this section, we will see the unique advantage making the ODWS basis superior to the recipe in Refs.~\cite{Zhang1988PLB,Zhang1991NPA}.

%----------------------------------------------------------------------------------------
\begin{figure}[htbp]
  \centering
  \includegraphics[width=0.8\textwidth]{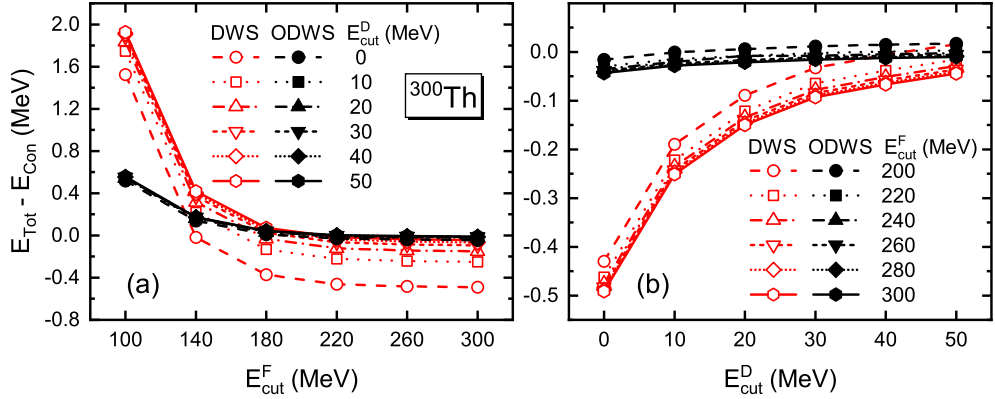}
  \caption{The difference of the total energy from the converged result as a function of $E_{\mathrm{cut}}^F$ in (a) and of $E_{\mathrm{cut}}^D$ in (b) for $^{300}$Th. The results from $E_{\mathrm{cut}}^D = 0, 10, \cdots, 50$ MeV in (a) and $E_{\mathrm{cut}}^F = 200, 220, \cdots, 300$ MeV in (b) are given by different symbols. Empty and filled symbols represent the results with the DWS and ODWS basis, respectively.}
\label{fig4}
\end{figure}
%----------------------------------------------------------------------------------------

To test the performance of the ODWS basis with respect to the convergence behavior, Fig.~\ref{fig4}(a) shows the difference of the total energy from the converged result as a function of $E_{\mathrm{cut}}^F$ in both calculations with the DWS and ODWS bases for $^{300}$Th.
It is easy to see that although both calculations lead to convergence with increasing cutoff energies, the results obtained with the ODWS basis are always closer to the converged one.
In particular, when $E_{\mathrm{cut}}^F = 100$ MeV, the total energy calculated with the ODWS basis is $\ge 1$ MeV lower than that with the DWS basis.
The results with the ODWS basis not only exhibit better convergence with the increasing $E_{\mathrm{cut}}^F$, but also hardly change with the increasing $E_{\mathrm{cut}}^D$.
This suggests that we can obtain converged results even without the bases from continuum in the Dirac sea, which would significantly reduce the basis space.

To more finely evaluate the convergence with respect to $E_{\mathrm{cut}}^D$, in Fig.~\ref{fig4}(b) we reduce the range of $E_{\mathrm{cut}}^F$ from [100, 300] MeV in Fig.~\ref{fig4}(a) to [200, 300] MeV, and thus we rescale the vertical axis.
It is found that for the usual DWS basis, the calculated total energy changes by about 0.5 MeV from $E_{\mathrm{cut}}^D = 0$ to 50 MeV, suggesting the importance of the bases from continuum in the Dirac sea.
In contrast, for the ODWS basis, the calculated total energy is almost a constant under different cutoff energies.
For example, the difference between the results from $E_{\mathrm{cut}}^F = 200$ MeV, $E_{\mathrm{cut}}^D = 0$ MeV and $E_{\mathrm{cut}}^F = 300$ MeV, $E_{\mathrm{cut}}^D = 50$ MeV is only 0.005 MeV, about $10^{-6}$ of its total energy.
Therefore, the calculation using the ODWS basis can provide converged total energies for heavy nuclei such as $^{300}$Th, with $E_{\mathrm{cut}}^F = 200$ MeV and $E_{\mathrm{cut}}^D = 0$ MeV that are both smaller than those required for the DWS basis.
We have further reduced $E_{\mathrm{cut}}^D$ to negative values to see if a smaller ODWS basis space in the Dirac sea could give converged results.
It turns out that $E_{\mathrm{cut}}^D \lesssim -100$ MeV leads to a ground-state deformation completely different from the converged one, and the difference between the total energy from $E_{\mathrm{cut}}^D \approx -50$ MeV and the converged one is more than 0.3 MeV.
This indicates that some bound bases within 50 MeV from the continuum threshold in the Dirac sea are important for convergence.
Thus, an energy cutoff of $E_{\mathrm{cut}}^D = 0$ MeV, i.e., the truncation right at the continuum threshold in the Dirac sea, is suggested for the ODWS basis to guarantee convergence for different nuclei.

%----------------------------------------------------------------------------------------
\begin{figure}[htbp]
  \centering
  \includegraphics[width=0.4\textwidth]{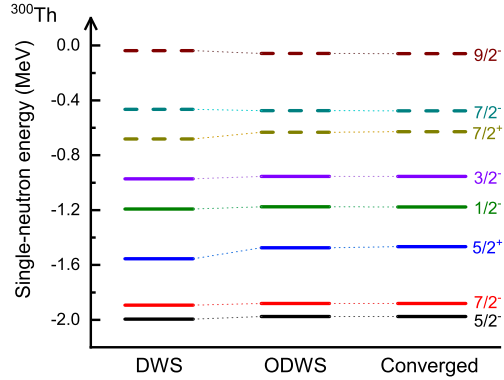}
  \caption{Single-neutron levels of $^{300}$Th near the continuum threshold calculated using DWS and ODWS basis with $E_{\mathrm{cut}}^F = 200$ MeV and $E_{\mathrm{cut}}^D = 0$ MeV as well as those from the converged result. The quantum numbers $m^\pi$ are given on the right, and the solid and dashed lines represent occupied and unoccupied levels, respectively.}
\label{fig5}
\end{figure}
%----------------------------------------------------------------------------------------

It is also interesting to check whether the ODWS basis space truncated by $E_{\mathrm{cut}}^F = 200$ MeV and $E_{\mathrm{cut}}^D = 0$ MeV can guarantee the convergence of single-particle properties.
Figure~\ref{fig5} compares the single-neutron levels of $^{300}$Th near the continuum threshold calculated using DWS and ODWS bases as well as those from the converged result.
Distinguishable differences can be seen between the result from the DWS basis and the other two.
For example, the difference for the single-neutron energy of the $5/2^+$ state is nearly 0.1 MeV.
On the other hand, the result from the ODWS basis and the converged one are almost the same, with the largest difference less than 0.01 MeV.
Therefore, according to this comparison and the convergence of the total energy in Fig.~\ref{fig4}, $E_{\mathrm{cut}}^F = 200$ MeV and $E_{\mathrm{cut}}^D = 0$ MeV for the ODWS basis are enough to give converged results in the DRHBc calculation.
They correspond to a certainly smaller basis space than that of the DWS basis and, thus, require less computing resource.

Finally, one may argue that using the converged mean-field potentials to construct the ODWS basis makes little sense, since one never knows \emph{a priori} the converged potentials of the calculated nucleus.
This is not true.
A natural way out is to construct a new basis set with new potentials in each step of the iterative solution of RHB equations.
This has been done and can lead to the same results as those calculated by the ODWS basis, but it is found that the reconstruction of the basis in each step is time consuming.
In practice, we find that during the iteration only one reconstruction of the basis is enough to get the same results.
Here the reconstruction happens when a reference parameter $\sigma \approx 1$ MeV, in which $\sigma$ means the largest difference of the mean-field potentials between the present step and the previous step during the iteration.
(When $\sigma$ is less than a certain value, e.g., $10^{-4}$ MeV, the self-consistent iteration converges.)
Note that there is no need to change the cutoff energies during the iteration in the above approach; namely, $E_{\mathrm{cut}}^F = 200$ MeV and $E_{\mathrm{cut}}^D = 0$ MeV are used for the original DWS basis from the beginning and for the ODWS basis after $\sigma \approx 1$ MeV, up to convergence.
It is also worthwhile to mention that, although this approach replaces bases during the iteration, there is almost no difference in the convergence steps compared with the calculation using only the DWS basis.
Therefore, the ODWS basis is practicable, and its application to large-scale DRHBc mass table calculations is expected.

%%%%%%%%%%%%%%%%%%%%%%%%%%%%%%%%%%%%%%%%%%%%%%%%%%%%%%%%%%
%                    begin  summary
%%%%%%%%%%%%%%%%%%%%%%%%%%%%%%%%%%%%%%%%%%%%%%%%%%%%%%%%%%

%----------------------------------------------------------------------------------------
\section{Summary}\label{summary}
%----------------------------------------------------------------------------------------

In summary, based on the deformed relativistic Hartree-Bogoliubov theory in continuum, we check in detail the convergence with respect to the basis space in the Dirac sea, and propose an optimized Woods-Saxon basis for nuclear density functional theory.
By checking the convergence with respect to the energy cutoff, we demonstrate that the basis space in the Dirac sea is not necessarily as large as that in the Fermi sea in order to obtain results with a satisfactory convergence accuracy of $10^{-5}$.
Then we use the converged mean-field potential to generate a new basis set, referred to as the optimized Dirac Woods-Saxon basis.
It is found that the new basis can lead to convergence in both total energy and single-particle levels with a basis space truncated by the energy cutoff $E_{\mathrm{cut}}^F = 200$ MeV in the Fermi sea and $E_{\mathrm{cut}}^D = 0$ MeV in the Dirac sea, which is substantially smaller than the original basis space required for convergence.
Finally, we suggest a method to construct the optimized Dirac Woods-Saxon basis during the iterative solution of relativistic Hartree-Bogoliubov equations, which guarantees the practicability of the new basis with a suitable basis space.

The application of the optimized Dirac Woods-Saxon basis to large-scale mass-table-type calculations based on the covariant density functional theory is expected to significantly reduce computing resource.
Similarly, its counterpart, the optimized Schr\"{o}dinger Woods-Saxon basis, is also available for nonrelativistic density functional theory.

\begin{acknowledgments}

Helpful discussions with members of the DRHBc Mass Table Collaboration are highly appreciated. This work was partly supported by the National Natural Science Foundation of China (Grants No.~11935003, No.~11875075, No.~11975031, and No.~12070131001), the National Key R\&D Program of China (Contracts No.~2017YFE0116700 and No.~2018YFA0404400), the State Key Laboratory of Nuclear Physics and Technology, Peking University (Grant No.~NPT2020ZZ01), and High-performance Computing Platform of Peking University.

\end{acknowledgments}

%merlin.mbs apsrev4-1.bst 2010-07-25 4.21a (PWD, AO, DPC) hacked
%Control: key (0)
%Control: author (72) initials jnrlst
%Control: editor formatted (1) identically to author
%Control: production of article title (-1) disabled
%Control: page (0) single
%Control: year (1) truncated
%Control: production of eprint (0) enabled
%

%======================================================================================%

\end{CJK*}
\end{document}